\documentclass[twocolumn]{aastex631}

\usepackage{xspace}

\usepackage{subfigure}

\pdfoutput=1 
\usepackage{amsmath}
\usepackage[T1]{fontenc}
\usepackage{url}
\usepackage{graphicx}
\usepackage{lipsum} 
\usepackage{float}
\usepackage{appendix}
\newcommand{\co}{\ensuremath{\mathrm{^{12}CO / ^{13}CO}\xspace}}

\newcommand{\Teff}{\ensuremath{T_\mathrm{eff}}\xspace}

\newcommand{\Rj}{\ensuremath{R_{\rm{Jup}}}\xspace}
\newcommand{\Mj}{\ensuremath{M_{\rm{Jup}}}\xspace} 
\usepackage{bm}

\begin{document}

\title{Chemical Links between a Young M-type T Tauri Star and its Substellar Companion: Spectral Analysis and  C/O Measurement of DH Tau A}

\author[0000-0001-5541-6087]{Neda Hejazi}
\affil{Department of Physics and Astronomy, University of Kansas, Lawrence,  KS 66045, USA}
\affil{Department of Physics and Astronomy, Georgia State University, Atlanta, GA 30303, USA}
\email{nhejazi@ku.edu}

\author[0000-0002-6618-1137]{Jerry W. Xuan}
\affil{Department of Astronomy, California Institute of Technology, Pasadena, CA 91125, USA}

\author[0000-0002-1221-5346]{David R. Coria}
\affil{Department of Physics and Astronomy, University of Kansas, Lawrence,  KS 66045, USA}

\author[0000-0002-8378-1062]{Erica Sawczynec}
\affil{Department of Astronomy, University of Texas, Austin, TX 78712 , USA}

\author{Ian J. M. Crossfield}
\affil{Department of Physics and Astronomy, University of Kansas, Lawrence,  KS 66045, USA}

\author[0000-0003-4019-0630]{Paul I. Cristofari}
\affil{Center for Astrophysics, Harvard \& Smithsonian, Cambridge, MA 02138, USA}

\author[0000-0002-3726-4881]{Zhoujian Zhang}\thanks{NASA Sagan Fellow}
\affil{Department of Astronomy and Astrophysics, University of California, Santa Cruz, CA 95064, USA}

\author{Maleah  Rhem}
\affil{Department of Physics and Astronomy, University of Kansas, Lawrence,  KS 66045, USA}

\begin{abstract}
The chemical abundance measurements of host stars and their substellar companions provide a powerful tool to trace the formation mechanism of the planetary systems. We present a detailed high-resolution spectroscopic analysis of a young M-type star, DH Tau A, which is located in the Taurus molecular cloud belonging to the Taurus-Auriga star-forming region.  This star is host to a low-mass companion, DH Tau b,  and both star and the companion are  still in their accreting phase. We apply our  technique \citep{Hejazi2024} to measure the abundances of carbon and oxygen using carbon- and oxygen-bearing molecules, such as CO and OH, respectively. We determine a near-solar carbon-to-oxygen abundance ratio of C/O=$\rm0.555\pm0.063$ for the host star DH Tau A. We compare this stellar abundance ratio with that of the companion from our previous study ($\rm C/O=0.54^{+0.06}_{-0.05}$, \citealt{Xuan2024b}), which also has a near-solar value.  This confirms the chemical homogeneity in the DH Tau system, which suggests a formation scenario for the companion consistent with a  direct and relatively fast gravitational collapse, rather than a slow core accretion process. 
\end{abstract}


\section{Introduction}\label{sec:intro}
Young, directly imaged planets and brown dwarf companions provide an excellent laboratory to study the initial conditions of planetary and stellar systems. In the past few years, the growing capabilities of high-contrast, high-resolution spectroscopy have provided us with increasingly robust atmospheric abundance measurements for dozens of substellar companions \citep[e.g.][]{Xuan2022, Wang2023, Zhang2024, Hsu2024, Landman2024, Xuan2024b}. The James Webb Space Telescope (JWST) is also ushering in an era of precise atmospheric measurements for substellar companions \citep{Miles2023, Gandhi2023} and isolated brown dwarfs \citep{Hood2024}.

The atmospheric compositions of planets, particularly the abundances of volatile elements, provide a fingerprint into their formation histories. Because of their influence on exoplanet ice and gas chemistry, the stellar abundances of volatile elements like H, C, N, O, S, and possibly isotopologue ratios such as $^{12}$C/$^{13}$C make excellent planetary formation and evolution diagnostics \citep[e.g.][]{Fortney_2012, Turrini_2021, Crossfield_2023, Ohno2023, Zhang_2021a, Zhang_2021b, Chachan2023, Coria_2024}. To the first order, host-companion carbon and oxygen abundance ratio comparisons may be used to distinguish between brown dwarfs, which form top-down, vs. gas giant planets, which form bottom-up. Brown dwarfs form via gravitational instability \citep[e.g.][]{offner_Formation_2010, bate_stellar_2012, Kratter2016, Ilee_2017, Hawkins_2020}, a much faster (sub-Myr timescales) process than core accretion \citep{pollack_formation_1996}. Giant planets form much slower via core accretion on Myr-timescales which allow protoplanets to incorporate varying quantities of gas and solids into their atmospheres, potentially resulting in a wide range of atmospheric metallicities, C/O ratios, and $^{12}$C/$^{13}$C ratios \citep[e.g.][]{pollack_formation_1996, Oberg2011, Alibert2013, Alibert2017, Bergin2024}.

Several studies have already identified emerging trends that could delineate different formation pathways. For example, \citet{Xuan2024b} showed that a sample of eight widely-separated, $\approx10-30~\Mj$ companions have broadly solar carbon and oxygen abundances, which could point towards these companions forming via top-down gravitational collapse and representing the tail end of binary star formation (see also \citealt{Hoch2023}). This conclusion depends on the carbon and oxygen compositions of the host stars, which \citet{Xuan2024b} argued to most likely be solar given the measured solar abundances of other stars in the same star forming regions \citep[e.g.][]{Santos2008, DOrazi2011, Biazzo2017}. In contrast, giant planets with $m\lesssim10~\Mj$ appear to have atmospheres that are enriched in metals compared to their stars \citep[]{molliere_Retrieving_2020, Wang2023, Zhang2023, Nasedkin2024}. 

As we obtain improved compositional measurements for imaged planets and brown dwarfs, it is crucial to also advance our knowledge of their host star abundances. The exoplanet systems most amenable to direct host-companion abundance comparisons typically consist of a well-characterized, field-age, Sun-like star; however, a large subset of host stars to directly imaged companions have late K or M spectral types. Traditional stellar abundance measurements calibrated to work with optical data are often insufficient for these cooler stars because of the overwhelming metal absorption lines present in their spectra. This means that spectral information on their carbon and oxygen content is most accessible using molecular absorption features in the near-infrared (NIR).  There are several studies that successfully measure elemental abundances for older K/M dwarf stars \citep[e.g.][]{Souto_2017, Souto_2018, Souto_2022, Hejazi2023, Hejazi2024}; however, there are numerous challenges involved deriving abundances for host stars of directly imaged companions, which are predominantly young ($\lesssim100$ Myr). For example, rapid rotation of young stars causes significant line blending \citep[e.g.][]{Zhang2023}, magnetic fields alter the line profiles of many atomic lines \citep[e.g.][]{Cristofari2023_SPIROU}, and veiling effects complicate abundance measurements from absorption lines by adding an extra layer of continuum emission \citep[e.g.][]{LopezValdivia2021}. In this paper, we present a near-infrared spectral analysis for one such young, challenging, cool dwarf star host to a high-priority directly imaged companion: DH Tau A.

DH Tau A is a young, accreting M2.3 dwarf \citep{Roccatagliata2020} located at a distance of 133.4 pc \citep{GAIA_EDR3} from the Sun. DH Tau A forms an ultrawide binary system with DI Tau at a projected separation of 15$\arcsec$ \citep[e.g.][] {Kraus_Hillenbrand_2009}. DH Tau is also host to a possibly accreting \citep{zhou_2014, bonnefoy_2014, Holstein2021, Martinez_Kraus2021}, widely separated ($\approx310$ AU), low-mass companion DH Tau Ab  (or for simplicity, DH Tau b), which was first discovered by \cite{Itoh2005}. Previous SPHERE observations related to immediate surroundings of substellar objects including DH Tau b hinted at the existence of a point source close to DH Tau b with an estimated mass of $\approx1~\Mj$, which if confirmed, it would be the first of its kind, leading to profound insights into the formation, evolution, and occurrence of such giant pairs. Recently, \citet{Xuan2024b} has inferred a mass of $12\pm4~\Mj$ for DH Tau b based on its bolometric luminosity and a system age of $0.7^{+0.3}_{-0.1}$ Myr. Using $K$ band high-resolution spectra from $2.29-2.49~\mu$m collected by Keck/KPIC, which show clear CO and H$_2$O absorption lines in the companion, \citet{Xuan2024b} also performed atmospheric retrieval analyses using \texttt{petitRADTRANS} \citep{molliere_Retrieving_2020} to constrain  C and O abundances, and measured a near-solar $\rm C/O=0.54^{+0.06}_{-0.05}$ ratio and $\rm [C/H]=-0.32^{+0.34}_{-0.30}$ for DH Tau b.  Here, we measure the C/O ratio of the host star DH Tau A to provide a direct comparison to that of DH Tau b. To do so, we update the methodology presented in \citet{Hejazi2024} to account for the effects of rotation and veiling.

 This paper is organized as follows. In Section \ref{sec:obs}, we summarize the spectroscopic observations of the host star DH Tau A and the data reduction technique as well as  the pre-processing needed to prepare the spectra for the abundance analysis. The physical parameter determination of the host star is presented in  Section \ref{sec:star_parameters}. The measurements of the elemental C and O abundances and C/O abundance ratio of the host star along with their error analysis are detailed in Section \ref{sec:star_abundance_analysis}.  We discuss the chemical connection between the host star and companion in DH Tau system in Section \ref{sec:discussion}. The summary of this study is presented in Section \ref{sec:summary}.

\section{Observations}\label{sec:obs}
The Immersion GRating Infared Spectrograph (IGRINS; \citealt{Yuk2010, Wang2010, Gully-Santiago2012, Han2012, Moon2012, Jeong2014, Oh2014, Park2014}) is a cross-dispersed high-resolution (R$\sim$45,000) spectrograph providing simultaneous spectra in both the $H$ and $K$ band (1.45 to 2.48 $\mu$m) in a single exposure.
IGRINS observations of DH Tau A were taken on November 6, 2015 using the 2.7m Harlan J. Smith Telescope at McDonald Observatory \citep{Mace2016} and obtained using the public Raw \& Reduced IGRINS Spectral Archive (\href{https://igrinscontact.github.io/}{RRISA}; \citealt{Sawczynec2023}; 2024, in preparation).
The observations of DH Tau A were paired with observations of a nearby A0V standard star, k Tau, to allow for correction of telluric features in the spectra. Both DH Tau A and k Tau were observed using a single ABBA nod sequence along the slit, using a standard slit position angle of 90$^{\circ}$ east of north which avoids flux contamination from nearby companion DH Tau b for the science spectra. Each individual frame for DH Tau A had an exposure time of 250 seconds resulting in a 1D reduced spectral signal-to-noise ratio (SNR) of $\sim$120 and $\sim$140 per resolution element in the $H$ and $K$ band, respectively\footnote{Through our spectroscopic analyses of cool dwarfs using IGRINS spectra, we have determined  a minimum SNR of $\sim$200 per resolution element required for simultaneously measuring the abundances of different elements with sufficient accuracy \citep{Hejazi2023, Hejazi2024}. Fortunately, OH lines are slightly affected by spectral noise, due to the numerous well-shaped lines that reside in the $H$ band \citep{Melo2024}. Prominent CO lines  in the $K$ band, however, are relatively more influenced by noise, which is why we have identified fewer CO lines appropriate for this analysis (see Table \ref{tab:line_data}).}

The observations were reduced using a beta version of the IGRINS Pipeline Package (\href{https://github.com/igrins/plp}{\texttt{IGRINS PLP v3}}; \citealt{Kaplan2024}), which performs standard echelle spectroscopy reduction techniques tuned for IGRINS data. Data reduced using the \texttt{IGRINS PLP v3} is cosmic ray and instrumental flexure corrected (Sawczynec et al. 2024, in preparation) before the individual exposures for each slit position (A/B) are stacked. The stacked A nod is then subtracted from the stacked B nod, removing any background contributions from the sky, thermal emission, stray light, or hot pixels. The detector readout pattern is removed before the echellogram is flat fielded and the individual 2D echelle orders are rectified. The 1D spectra are generated using a modified version of the optimal extraction described in \citet{Horn1986} for each order. The wavelength solution for the spectra are calculated using 2D polynomial fitting of the known OH emission lines locations in the echellogram of the sky frames for the night to convert detector position to wavelength. Finally, the 1D reduced DH Tau A spectra were divided by the 1D reduced k Tau spectra (airmass difference $\sim$0.004) and multiplied by a \href{http://kurucz.harvard.edu/stars.html}{model of Vega} from \citet{Kurucz1979} to produce the final relative flux calibrated and telluric corrected science spectra for DH Tau A.

We further processed the spectra using the \texttt{SpeXTool} pipeline's \texttt{xtellcor\_basic} routine \citep{Cushing2004} to account for any small wavelength offset between the spectra of DH Tau A and A0V standard star, and then using \texttt{xmergeorders} routine to trim out wavelength areas with large telluric residuals and combine the individual echelle orders into a single, 1D spectrum. We then flattened the observed 1D spectra using the method described in \cite{Hejazi2023, Hejazi2024}. We note that the resulting flattened spectra do not present continuum-normalized spectra, but these are (pre)processed spectra that will be used for continuum/pseudocontinuum normalization and abundance measurements (see Section \ref{sec:elemental_abundances}).

\section{Physical Parameters Of the Host Star DH Tau A}\label{sec:star_parameters}
Physical parameters of DH Tau A have been measured by various studies. However,  many of these analyses have not taken into account the effects of magnetic field and veiling that are of importance in the atmosphere of young stellar objects \citep[e.g.][]{LopezValdivia2021}.  The splitting of spectral lines into multiple components by magnetic field (Zeeman effect) can change the shape of spectral lines \citep[e.g.][]{Reiners_Basri_2007}. Veiling (R) is a nonstellar continuum emission as a result of several physical processes such as chromospheric activity \citep[e.g.][]{Calvet1984}, emission from accretion flow  onto the star or in the vicinity of the stellar surface \citep[e.g.][]{Kenyon_Hartmann1987}, and  emission from gas and dust in the surrounding disk \citep[e.g.][]{Natta2001, Fischer2011}. The effect of veiling emerges as an additional continuum superimposed on the stellar spectrum, which decreases the depth and, and in turn, the equivalent width of the spectral lines \citep[e.g.][]{Joy1949,Stempels_Piskunov_2003}. As a result, the physical parameters of young stars derived from methods that do not include these two effects may not be reliable. 

We find a significant degeneracy between veiling R and metallicity [M/H]; a decrease in [M/H] causes a decrease in the depth of spectral lines, which can be mostly compensated with a decrease in R, and vice versa. As a result, a spectrum may be well fit with multiple synthetic models having different pairs of [M/H] and R. Simultaneous variations of these two parameters during the model fitting process thus may give rise to a false best-fit model. To overcome this problem, we opt to use a  metallicity value  obtained from an independent study  and keep this parameter fixed in our synthetic spectral  fitting. However, there is no unique and reliable determined metallicity for DH Tau A in the literature among the reported metallicity values ranging from near solar metallicities ([M/H]$\simeq$0, e.g., \citealt{Jonsson2020, Sprague2022}) into the low-metallicity regime ([M/H]$\simeq$$-$1, e.g., \citealt{Abdurro'uf2022, Wang_2023}). Furthermore, current photometric metallicity relations have been calibrated using field (and mostly old) M dwarfs and may not be applied to young, accreting M-dwarf stars. Thus, we have employed photometric calibrations  of M-dwarf metallicity from \cite{Duque-Arribas2023} but found unrealistic values for  DH Tau A. We have also used the photometric calibrations from \cite{Mann2015, Mann2019} to derive the effective temperature as well as mass and radius that can be converted to surface gravity \citep[e.g.][]{Hejazi2022} and again found the parameter values for DH Tau A far  beyond the ranges valid for M-type stars. This prompted us to search for yet another metallicity proxy for DH Tau.

DH Tau A belongs to the Taurus-Auriga star-forming region that encompasses the Taurus molecular cloud containing hundreds of newly formed stars. Using seven low-mass members of the Taurus-Auriga region, including both classical T Tauri (similar to DH Tau A) and weak-lined stars, \cite{DOrazi2011} obtained a mean metallicity of [M/H]=$-$0.01$\pm$0.05 and the solar abundances for the $\alpha$ element Si and the Fe-peak element Ni. Accordingly, the derived metallicity values of DH Tau A found in  the literature that are significantly different from the solar metallicity  likely suffer from large biases. For this reason, we assume a solar metallicity with a typical uncertainty of 0.10 dex \citep[e.g.][]{Hejazi2024} for DH Tau A. In Section \ref{sec:abundance_errors}, we explain that even a larger uncertainty for metallicity (as well as other physical parameters) would not change the uncertainty of C/O ratio noticeably.

We obtain the physical parameters of DH Tau A using the method described in \cite{Cristofari2023_SPIROU}, where a new code  \texttt{ZeeTurbo} was introduced by incorporating the Zeeman effect and polarized radiative transfer capabilities to the widely-used radiative transfer code \texttt{Turbospectrum} \citep{AlvarezPlez1998, Plez2012}. We  implement a Markov chain Monte Carlo (MCMC) analysis based on  the emcee package \citep{Foreman-Mackey2013} to simultaneously estimate the physical parameters along with the average surface magnetic flux as well as their corresponding uncertainties for DH Tau A as follows: effective temperature $\Teff = 3726~\pm$ 30 K, surface gravity log$(g) = 4.00~\pm$ 0.05 dex, projected rotational velocity V$_{\rm rot}\sin \textit{i} = 7.1~\pm$ 0.1 km s$^{-1}$ (which is higher than typical rotational velocities of old M dwarfs), veiling in $H$ band R$_{H} = 0.90~\pm$ 0.03, veiling in $K$ band R$_{K} = 0.99~\pm$ 0.03\footnote{We treat the veiling parameter following Eq. 1 in \cite{LopezValdivia2021}, which was first suggested by \cite{Basri_Batalha1990}. Two best-fit veiling values have been determined for all the wavelengths in $H$ band and $K$ band, separately.}, and the average magnetic field B = 2.8 $\pm$ 0.1 kG. {The errors bars are estimated from the posterior distributions and account for both photon noise and  systematics, following the detailed approach from \cite{Cristofari2022, Cristofari2023_SPIROU}.  We use these parameters as input in our abundance analysis. The inclusion of  magnetic field in the MCMC process certainly results in more accurate values for other parameters, in general. However, we find that the molecular OH and CO lines, that are used to measure the O and C abundances, respectively (see Section \ref{sec:star_abundance_analysis}), are not considerably sensitive to magnetic field, and the magnetic effect on these lines can be ignored. For this reason, we do not take this effect into account for our abundance measurements. The physical parameters of DH Tau A inferred from several studies are listed in Table \ref{tab:parameters} for comparison. By visual inspection, we find the synthetic model associated with the parameters obtained from this study shows the best match with the observed spectrum, as compared to the models constructed using the parameters derived from other studies.

\begin{deluxetable*}{cccccc}
\tablecaption{The physical parameters and abundance ratios of DH Tau A (Host) vs. DH Tau b (Companion) }
\tablewidth{0pt}
\tablehead{
\colhead{Host} & \colhead{\cite{LopezValdivia2021}} & \colhead{\cite{Abdurro'uf2022}} & \colhead{\cite{Yu2023}} & \colhead{\cite{Wang2024}} & \colhead{This work} 
}
\startdata
$\Teff$ (K) &  3477$\pm$125 & 3594.7$\pm$7.8 & 3594.70$\pm$106.00 & 3751.50$\pm$... & 3726$\pm$30\\
log g   &   3.89$\pm$0.22 & 3.560$\pm$0.026 & 3.560$\pm$0.100 & 4.00102997$\pm$... & 4.00$\pm$0.05 \\
$\rm{V_{rot}}\sin \textit{i}$ (km s$^{-1}$) &  8.4$\pm$2.1 & 10.58766$\pm$... & ... & ... & 7.1$\pm$0.1 \\
B (kG) &  2.21$\pm$0.32 & ... & ... & ...  & 2.8$\pm$0.1  \\
C/O & ... & ...  & ... & ... & 0.555$\pm$0.063 \\
\hline
\hline
\colhead{Companion} & \colhead{\cite{Itoh2005}} & \colhead{\cite{Patience2012}} & \colhead{\cite{zhou_2014}} & \colhead{\cite{vanHolstein2021}} & \colhead{\cite{Xuan2024b}}\\
\hline
Mass ($\Mj$) & 30-50 & $11^{+10}_{-3}$ & 11$\pm$3 & $15^{+7}_{-4}$ & $12\pm4$ \\
Radius ($\Rj$) & ... & ... &  2.7$\pm$0.8 & ... & $2.6\pm0.6$ \\
$\Teff$ (K) & 2700-2800 & 2350$\pm$150 & 2200$\pm...$ & $2400\pm100$ & $2050^{+120}_{-100}$ \\
$\rm{V_{rot}}\sin \textit{i}$ (km s$^{-1}$) & ... & ... & ... & ... & $5.7^{+0.8}_{-1.0}$ \\
C/O & ... & ... & ... & ... & $0.54^{+0.06}_{-0.05}$ \\
$^{12}$C/$^{13}$C & ... & ... & ... & ... & $53^{+50}_{-23}$ \\
\enddata
\tablecomments{The companion parameters reported in this table from \citet{Xuan2024b} are from atmospheres retrievals assuming a clear atmosphere. \citet{Xuan2024b} noted that the evolutionary models predict a higher $\Teff$ of $2350\pm200~$K instead. No number has been shown if there is not any  inferred value for a specific parameter from a particular paper.}
\label{tab:parameters}
\end{deluxetable*}

\section{Chemical Abundance Analysis of the Host Star DH Tau A}\label{sec:star_abundance_analysis}
\subsection{Elemental Abundance Measurements}\label{sec:elemental_abundances}
The atomic lines of light elements such as O and C are too weak and blended to be identified in the NIR spectra of M dwarfs. On the other hand, there are a significant number of prominent OH lines (in the $H$ band) and CO lines (in the $K$ band), which can be used to measure the abundances of these two elements. We therefore measure the elemental carbon and oxygen abundances using molecular CO and OH lines, respectively, through a self-consistent, iterative approach. Our abundance analysis is based on the method described in \cite{Hejazi2023, Hejazi2024} using an automatic code, ``\texttt{AutoSpecFit}'', which, in conjunction with the Turbospectrum code \citep{AlvarezPlez1998,Plez2012}, MARCS model atmospheres \citep{Gustafsson2008} and a set of atomic and molecular line lists,  carries out an iterative, line-by-line ${\chi}^{2}$ minimization over a set of selected spectral lines to measure the abundances of individual elements, simultaneously.

We identify the OH and CO lines that are nearly isolated from other species and have a well-defined shape (e.g., not distorted by noise or bad pixels, for example, due to artifacts or incomplete data reduction) and are also strong enough to be distinguished from the prevalent background H$_{2}$O opacities common in M-dwarf spectra. In order to select the best lines for our study, the observed spectrum is first normalized relative to an initial guess of best-fit model over each line candidate. We generate a synthetic model (hereafter, Mod{$_{\textrm{app}}$) associated with the star's physical parameters (Section \ref{sec:star_parameters}) and the solar absolute abundances for all elements\footnote{An estimate of the star's best-fit model is associated with an absolute abundance equal to the solar absolute abundance plus metallicity [M/H], i.e., {A(X)=A(X)$_{\sun}$+[M/H]}, for each element X, where A(X) is absolute abundance (Eq. \ref{equ:abs_abund_def}). However, for this study, the metallicity of DH Tau A is assumed to be zero.}, which is an approximation of the star's best-fit model.  The most appropriate ``normalizing'' data points within the neighboring continuum/pseudocontinuum regions around each line candidate are specified through a linear fit to the residual, R=obs/syn, as a function of wavelength, where ``obs'' is the observed flux and ``syn'' is the synthetic flux, usually along with two iterative {$\sigma$}-clippings (2{$\sigma$} and 1.5{$\sigma$}). This requires a trial-and-error inspection where different wavelength intervals in the continuum/pseudocontinuum around the line of interest are examined until at least one or two data points on each side of line are determined. On occasion, a couple of lines are close to one another, and the same normalizing regions  around,  and  in some cases, also between these lines are chosen. The observed spectrum over each specific line is normalized after dividing the spectrum by the linear fit to the residuals at the determined normalizing data points. In Figures \ref{fig:OH_normalization} and \ref{fig:CO_normalization}, we illustrate the selected normalizing regions  and the corresponding normalizing data points for  OH  and   CO lines, respectively. The blue line shows the synthetic model Mod{$_{\textrm{app}}$ and the red dots show the observed data normalized to this model using the normalizing points. The pseudocontinuum around spectral lines is dominated by prevailing, weak H$_{2}$O lines, many of which are not well modeled due to incomplete H$_{2}$O line lists. This generally results in a discrepancy between the observed and model spectrum within some pseudocontinuum regions (while these weak H$_{2}$O lines have a negligible effect on the prominent analyzed lines), which makes it challenging to find appropriate data points for normalization. For many of the analyzed lines, we have therefore been able to determine only a few normalizing data points on each side within narrow intervals in the pseudocontinuum, though are sufficient for the normalization process. We select the OH and CO lines that show a reasonable consistency between the normalized observed spectrum and the synthetic model Mod{$_{\textrm{app}}$}. We exclude any line for which there is a significant discrepancy in depth and/or shape between the normalized observed spectrum and the estimated best-fit model. Our selected lines, including 24 OH and 8 CO lines are listed in Table \ref{tab:line_data}. The oscillator strengths of these lines are also shown in the fourth column of this table. 

The above-determined normalizing continuum/pseudocontinuum regions corresponding to each selected line are recorded and will then be used as input in the next step. In addition, for each analyzed line, we manually select a fitting or ${\chi}^{2}$ window, mostly  far from the outermost part of the wings  as another input to perform the ${\chi}^{2}$ minimization process, which is shown in the third column of Table \ref{tab:line_data}. There are a variety of sources that may give rise to  noise in  stellar spectra such as photon noise or random fluctuations of the star light, stellar variability, sky background, the Earth’s atmospheric turbulence, and instrumental noise from telescope and detector. We find some spectral regions with relatively lower SNR values compared to other wavelength intervals, which are notably perturbed by noise,  resulting in a substantial mismatch between the observed and model spectrum. Nevertheless, even within these regions, there is still a good  agreement between the observed spectrum and synthetic model inside  most of these strong absorption lines around their cores. Our choice of fitting windows also considers the minor effect of the dominant, weak H$_{2}$O lines  on the inner region of the prominent analyzed lines. Together with the previously inferred  atmospheric parameters of the star (Section \ref{sec:star_parameters}), we run AutoSpecFit to measure the abundances of oxygen and carbon. We show our results in Table \ref{tab:results1};  the number of analyzed lines, \textit{N}, is shown in the second column, and the abundance [X/H] is presented in the third column, which is defined by

\begin{equation}\label{equ:abund_def}
\begin{split}
 {\rm [X/H]}_{\rm star} &=  {\rm log({N_{X}}/{N_{H}})_{star} -log({N_{X}}/{N_{H}})_{\sun}} \\ 
 &= {\rm A(X)_{star}-A(X)_{\sun}}
\end{split}
\end{equation}}

\noindent
where $\rm {N_{X}}$ denotes the number density of element X, $\rm {N_{H}}$ indicates the number density of hydrogen, and A(X) shows the absolute abundance, i.e.,

\begin{equation}\label{equ:abs_abund_def}
{\rm A(X)} =  {\rm log({N_{X}}/{N_{H}}) + 12}. \\ 
\end{equation}

\noindent
These results show  slightly supersolar carbon and oxygen abundances for DH Tau A.

Figures \ref{fig:OH_bestfit} and \ref{fig:CO_bestfit} show the comparison between the resulting best-fit synthetic model and the target's observed spectrum (normalized to the best-fit model) over all selected OH and CO lines, respectively. The best-fit model is associated with the star's physical parameters and the O and C abundances inferred from this analysis. The O and C abundances are the weighted average abundances of all the analyzed OH and CO lines, respectively (see \citealt{Hejazi2023, Hejazi2024} for more details), and the best-fit model does not necessarily show a perfect match to all these lines. Nevertheless, there is an excellent consistency between the observed spectrum and the determined best-fit model for most of the lines as seen in these figures by eye and confirmed by AutoSpecFit's ${\chi}^{2}$ minimization process.

\begin{figure}[h]
    \centering
    \includegraphics[width=1\linewidth]{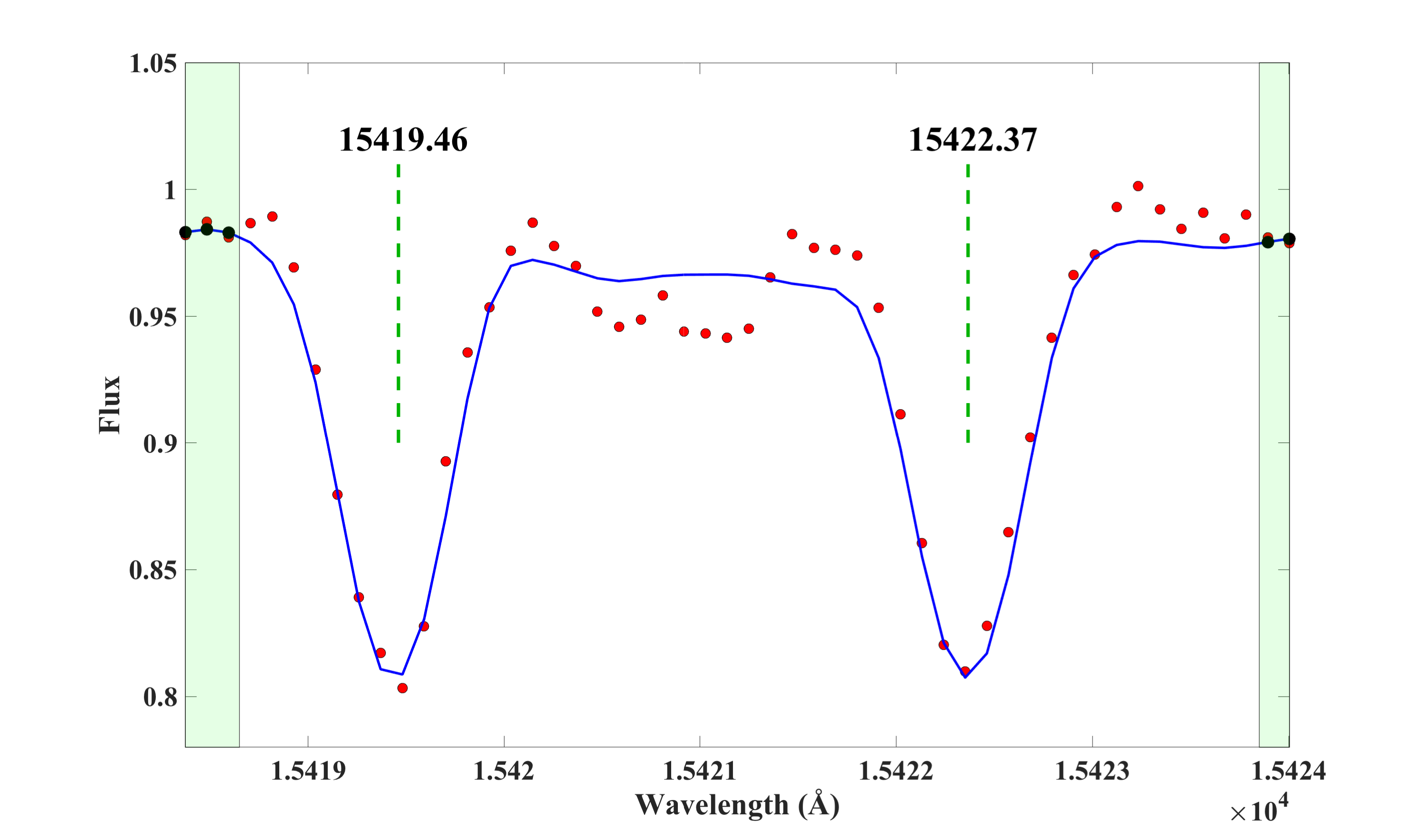}
    \caption{Comparison between an initial-guess of best-fit model Mod{$_{\textrm{app}}$} (blue line) and the observed spectrum of DH Tau A (red dots) normalized to this model over two adjacent OH lines using their common normalizing regions (green-shaded areas) and normalizing wavelength data points (black dots).} 
\label{fig:OH_normalization}
\end{figure}

\begin{figure}[h]
    \centering
    \includegraphics[width=1\linewidth]{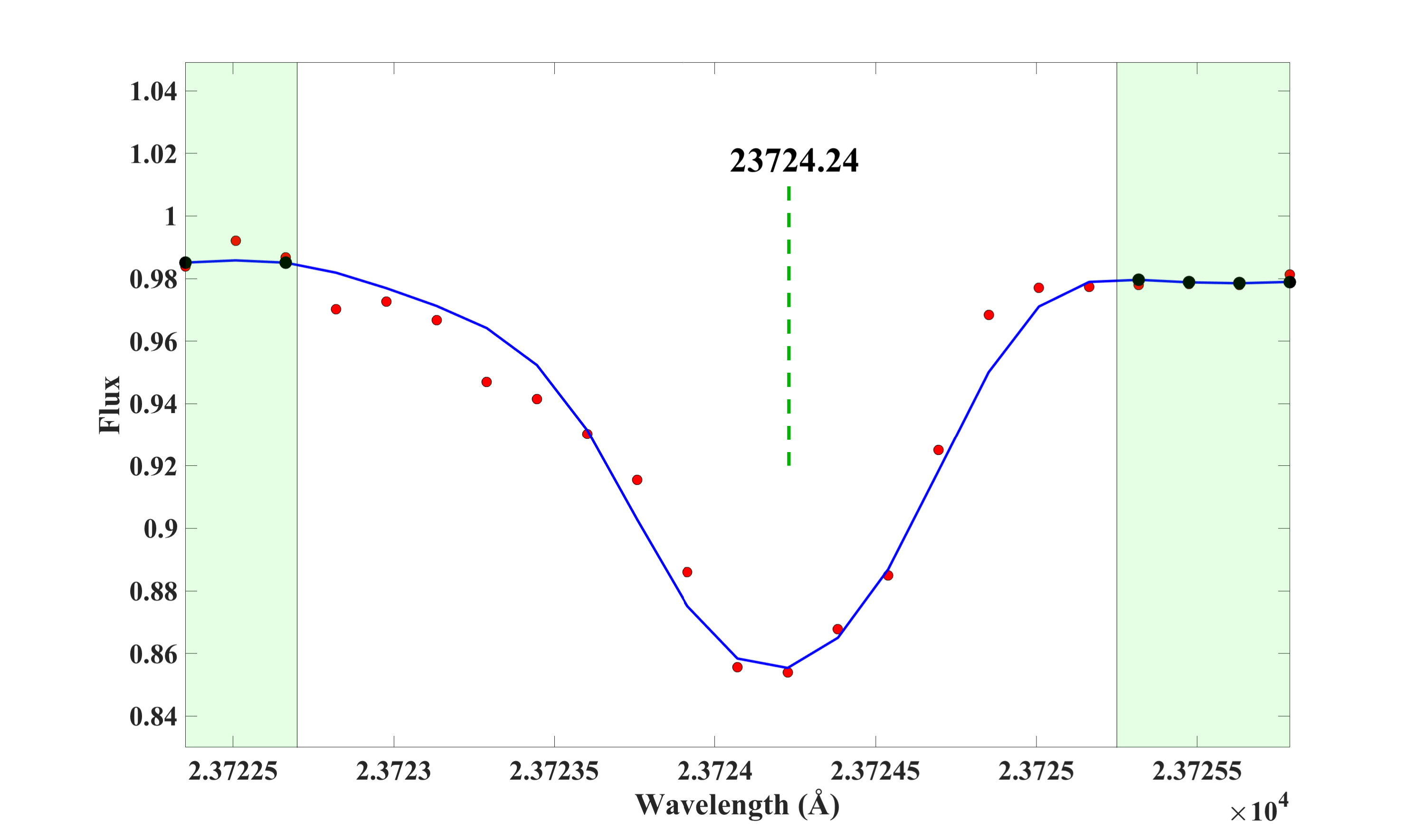}
    \caption{Comparison between an initial-guess of best-fit model Mod{$_{\textrm{app}}$} (blue line) and the observed spectrum of DH Tau A (red dots) normalized to this model over one single CO line using its respective normalizing regions (green-shaded areas) and normalizing wavelength data points (black dots).}
\label{fig:CO_normalization}
\end{figure}

\subsection{Abundance errors}\label{sec:abundance_errors}
To obtain the uncertainties of the inferred abundances, we first determine the random (statistical)  errors using the standard  error of the mean, i.e., $\rm{\sigma_{ran}}$=std/$\rm{\sqrt{N}}$, where std is the standard deviation of  the abundances from different lines of each species, as shown in the 9th column of Table \ref{tab:results2}. We also derive the systematic errors by determining the sensitivity of abundances to physical parameters. We deviate each parameter by its uncertainty (Section \ref{sec:star_parameters}) in both positive and negative direction one at a time (Tables  \ref{tab:results1} and  \ref{tab:results2}). We then carry out AutoSpecFit ten times, in each of which only one parameter is deviated in a specific direction while the other parameters are kept the same as the target's parameters, and the abundances of carbon and oxygen are inferred from each run. The columns 4-12 of Table \ref{tab:results1} and the columns 2-7 of Table \ref{tab:results2} show the abundance (or absolute abundance)\footnote{Based on Eq. \ref{equ:abund_def}, since the Solar absolute abundance, A(X)$_{\sun}$, is constant, $\rm{{\Delta}[X/H]_{star}={\Delta}A(X)_{star}}$.} variations due to the deviated parameters in both positive and negative directions relative to the inferred C and O abundances (i.e., [X/H], the third column of Table \ref{tab:results1})  as well as the average of absolute values of these variations ($\overline{\rm {\Delta}abund}$)  related to each parameter.   

The above variations, in general, depend on the specific deviated parameter and the direction of deviation, i.e., positive or negative, as well as the particular element, i.e., carbon or oxygen. For example, the oxygen abundance is more sensitive to [M/H], as compared to the carbon abundance. On the other hand, the carbon abundance is more sensitive to $\Teff$, log($g$), and V$\rm{_{rot}}\sin \textit{i}$, as compared to the oxygen abundance. The abundances of the two elements show roughly similar sensitivity to the veiling factor R with small differences depending on the  direction in which this parameter is deviated. The systematic abundance error, $\rm{\sigma_{sys}}$, of each element is determined by the quadrature sum of the $\overline{\rm {\Delta}abund}$ values from all parameters (the column 8 of Table \ref{tab:results2}):

\begin{figure*}[h]
    \centering
    \includegraphics[width=1\linewidth]{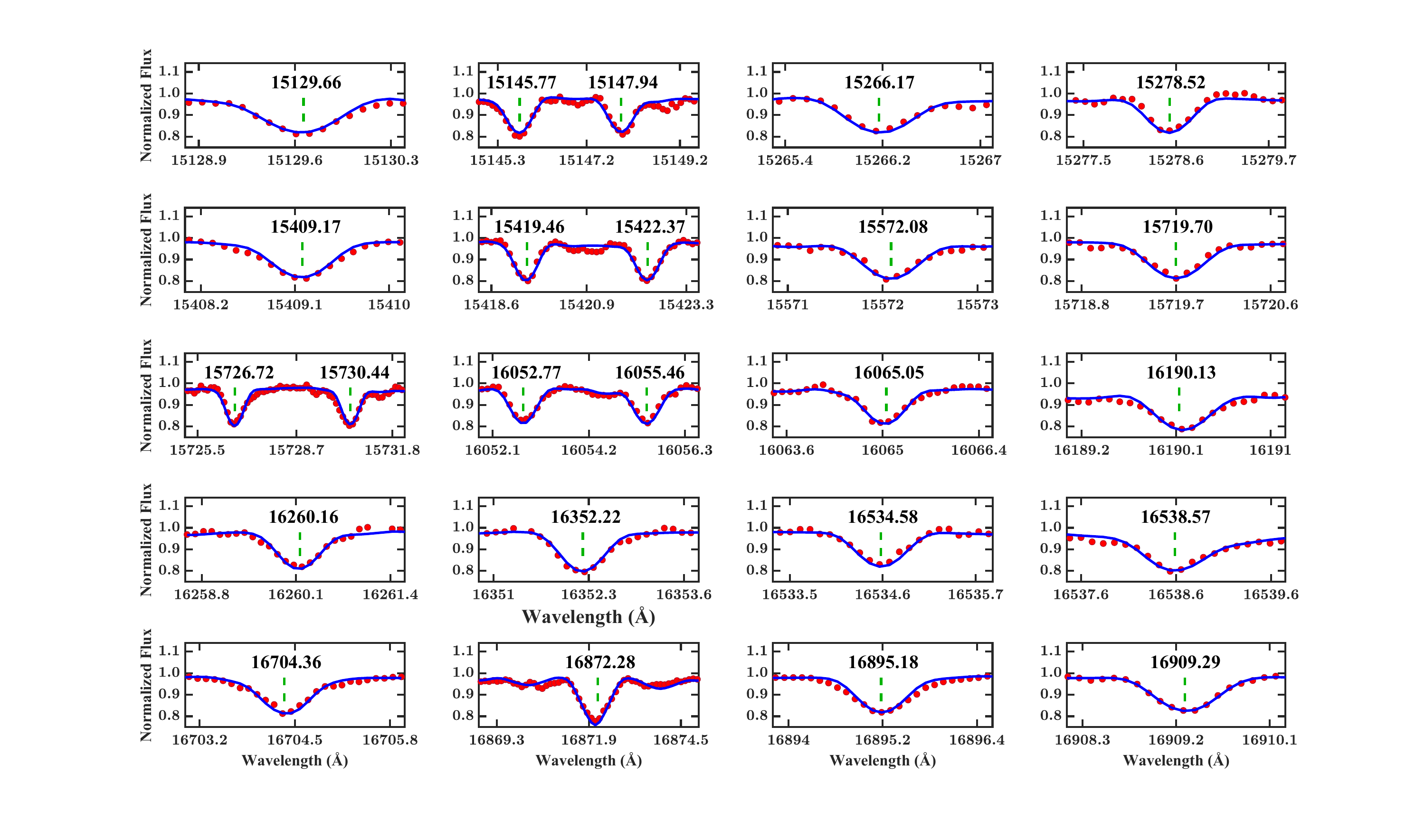}
    \caption{Comparison between the normalized, observed spectrum of DH Tau A (red dots) and the best-fit model (blue line) over the selected OH lines used to measure the oxygen abundance. The adjacent OH lines, which are normalized using common normalizing regions, are shown in the same panel. The location of the central wavelength of each OH line is shown by a green dashed line in the respective panel.}
\label{fig:OH_bestfit}
\end{figure*}

\begin{figure*}[h]
    \centering
    \includegraphics[width=1\linewidth]{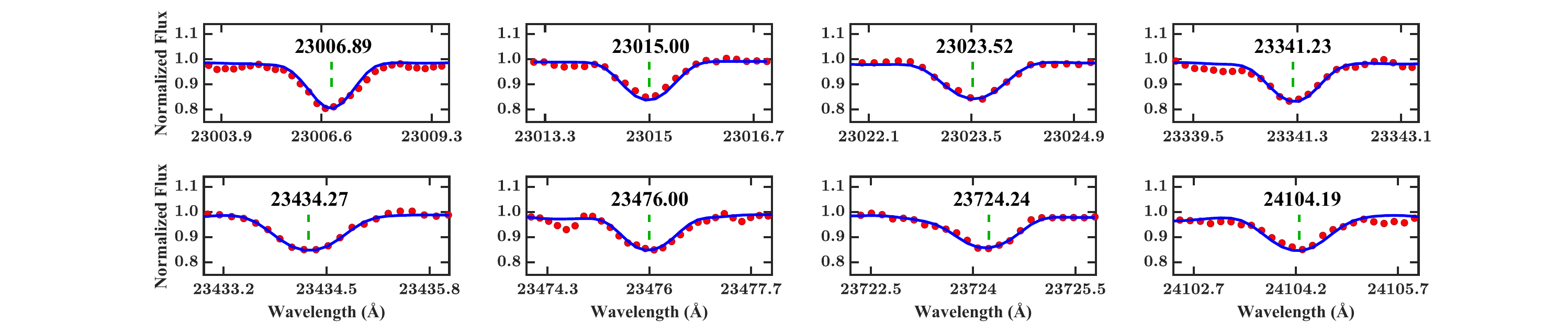}
    \caption{Comparison between the normalized, observed spectrum of DH Tau A (red dots) and the best-fit model (blue line) over the selected CO lines used to measure the carbon abundance. The location of the central wavelength of each CO line is shown by a green dashed line in the respective panel. }
\label{fig:CO_bestfit}
\end{figure*}

\begin{equation}\label{equ:sys_err}
{\rm \sigma_{sys}} = {\rm \sqrt{\sum_{{\textit{S}}}^{} [\overline{\rm ({\Delta}abund)_{\textit{S}}}]^2}} 
\end{equation}

\noindent
where the index \textit{S} takes  the sequence T, M, G, V, and R corresponding to the deviated parameters $\Teff$, [M/H], log($g$), V$_{\rm rot}\sin \textit{i}$, and R, respectively (Tables \ref{tab:results1} and \ref{tab:results2}). The total abundance error,  $\rm{\sigma_{tot}}$, is obtained by the quadrature sum of the random and systematic errors: 

\begin{equation}\label{equ:total_err}
\rm{\sigma_{tot}}=\sqrt{\sigma_{ran}^2 + \sigma_{sys}^2 }
\end{equation}

\noindent
as shown in the last column of Table \ref{tab:results2} for each element.

\subsection{Carbon-to-Oxygen Abundance Ratio}\label{sec:star_abundance_ratio}

We determine the star's C/O abundance ratio using the equation:

\begin{equation}\label{equ:abundance_ratio}
\rm{C/O=\frac{N_{C}}{N_{O}}=10^{(A(C)-A(O))}}
\end{equation}

\noindent
where $\rm {N_{C}}$ and $\rm {N_{O}}$ indicate the number densities and A(C) and A(O) indicate the absolute abundances of elements C and O, respectively. We find a nearly solar C/O=0.555 for DH Tau A. For reference, the solar ratio is (C/O)$_{\sun}$=0.540 based on the solar abundances from \citet{Grevesse2007} (which we also use for our abundance analysis, see \citealt{Hejazi2023, Hejazi2024} for additional information).

We employ the C and O abundance errors (Section \ref{sec:abundance_errors}) to calculate the uncertainty of the inferred C/O. Since the abundance ratio of the elements C and O depends on the subtraction of their absolute abundances (Eq. \ref{equ:abundance_ratio}), their  (correlated) systematic uncertainties related to the variation of each parameter largely cancel, making the contribution of all parameters (after adding in quadrature) to the total systematic error of C/O (Eq. \ref{equ:total_sys_err_abund_ratio}) relatively small. It should be noted that the increase of the parameter uncertainties (even by 100$\%$) would not significantly change the total error of C/O ratio. This is because the increase in the uncertainty of each parameter will change the abundances of C and O in the same direction (positive or negative) such that the change in the subtraction of A(C)-A(O) would remain small. In contrast, the (uncorrelated) random (statistical) errors of the two elements are added together in quadrature, though the overall contribution of the random error to the total uncertainty is also small. The systematic error of C/O ratio is calculated as follows. We first determine the difference in the $\overline{\rm {\Delta}abund}$ values between the two elements C and O for each parameter:

\begin{equation}\label{equ:abund_diff_C_O}
 {\rm \delta [A(C)-A(O)]}_{\textit{S}} = \overline{\rm ({\Delta}abund)_{\textit{S,C}}} -  \overline{\rm ({\Delta}abund)_{\textit{S,O}}} 
\end{equation}

\noindent
where \textit{S} indicates the same parameter sequence as described for Eq. \ref{equ:sys_err}. These values show how the variation of a given parameter would change the difference between the abundances (or absolute abundances) of carbon and oxygen. Following the derivative equation:

\begin{equation}\label{equ:derivative}
{\rm \delta(10^{x})} = {\rm ln(10)(10^{x})\delta(x)}
\end{equation}

\noindent
and using Eq. \ref{equ:abundance_ratio}, we derive the variation of C/O ratio due to the variation of ${\rm [A(C)-A(O)]}_{\textit{S}}$ for each parameter \textit{S} (Eq. \ref{equ:abund_diff_C_O}), as the systematic errors:

\begin{equation}\label{equ:sys_err_abund_ratio}
{\rm (\sigma_{sys})_{C/O,\textit{S}}} = {\rm ln(10)(C/O)\delta[A(C)-A(O)]}_{\textit{S}}
\end{equation}

\noindent
The total systematic error is obtained by the quadrature sum of all systematic errors:

\begin{equation}\label{equ:total_sys_err_abund_ratio}
{\rm (\sigma_{sys})_{C/O}} = {\rm \sqrt{\sum_{{\textit{S}}}^{} [(\sigma_{sys})_{C/O,\textit{S}}]^{2}}}
\end{equation}

\noindent
The random error is calculated in the same way:

\begin{equation}\label{equ:sys_err_abund_ratio}
{\rm (\sigma_{ran})_{C/O}} = {\rm ln(10)(C/O)({\delta}_{ran})_{C/O}}
\end{equation}

\noindent
but ${\rm ({\delta}_{ran})_{C/O}}$ is the  quadrature sum of the random errors of C and O abundances (the 9th column of Table \ref{tab:results2}).  Similar to Eq. \ref{equ:total_err}, we derive a total uncertainty of 0.063 for the C/O ratio of the host star DH Tau A.

\begin{deluxetable*}{lcccc}\label{tab:line_data} 
\tablecaption{The selected CO and OH lines used to measure the abundances of C and O of the host star DH Tau A}  
\tablewidth{0pt}
\tabletypesize{\scriptsize}
\tablehead{
\colhead{Species} &  Central wavelength ({\AA}) & {\ensuremath{{\chi}^2}} window ({\AA}) & log($gf$) & Comments}
\startdata
CO     & 23006.89 & 23005.80-23007.70 & $-$5.457 & Blended with one CO line: 23006.56, log($gf$)=$-$4.994  \\
CO     & 23015.00 & 23014.10-23015.85 & $-$5.474 & Blended with one CO line: 23014.87, log($gf$)=$-$4.985  \\
CO     & 23023.52 & 23022.75-23024.45 & $-$5.491 & Blended with one CO line: 23023.61, log($gf$)=$-$4.976  \\
CO     & 23341.23 & 23340.55-23342.00 & $-$5.065 & Blended with one CO line: 23341.14, log($gf$)=$-$4.481  \\ 
CO     & 23434.27 & 23433.45-23435.15 & $-$5.239 & \\
CO     & 23476.00 & 23475.20-23476.85 & $-$5.322 & Blended with two CO lines:  \\
       &          &                   &          & 23475.40, log($gf$)=$-$4.743 and 23476.80, log($gf$)=$-$5.075  \\
CO     & 23724.24 & 23722.90-23725.20 & $-$6.037 & Blended with one CO line: 23723.77, log($gf$)=$-$5.865  \\
CO     & 24104.19 & 24103.45-24105.10 & $-$5.456 & Blended with two CO lines:  \\ 
       &          &                   &          & 24104.25, log($gf$)=$-$4.590 and 24104.62, log($gf$)=$-$4.690  \\ 
\hline
OH     & 15129.66 & 15129.15-15130.15 & $-$5.499 & \\
OH     & 15145.77 & 15145.20-15146.25 & $-$5.447 & \\
OH     & 15147.94 & 15147.55-15148.40 & $-$5.447 & \\
OH     & 15266.17 & 15265.70-15266.65 & $-$5.429 & \\
OH     & 15278.52 & 15278.10-15278.95 & $-$5.382 & \\
OH     & 15409.17 & 15408.60-15409.70 & $-$5.365 & Blended with one OH line: 15409.09, log($gf$)=$-$6.605  \\
OH     & 15419.46 & 15418.85-15420.00 & $-$5.323 & \\
OH     & 15422.37 & 15421.85-15422.90 & $-$5.323 & \\
OH     & 15572.08 & 15571.60-15572.60 & $-$5.269 & \\
OH     & 15719.70 & 15719.20-15720.20 & $-$5.254 & \\
OH     & 15726.72 & 15726.20-15727.25 & $-$5.219 & \\
OH     & 15730.44 & 15729.90-15731.00 & $-$5.219 & \\
OH     & 16052.77 & 16052.30-16053.25 & $-$4.910 & \\
OH     & 16055.46 & 16054.95-16056.00 & $-$4.910 & \\
OH     & 16065.05 & 16064.50-16065.50 & $-$5.159 & \\
OH     & 16190.13 & 16189.70-16190.65 & $-$4.893 & Blended with one OH line: 16190.26, log($gf$)=$-$5.145  \\
OH     & 16260.16 & 16259.65-16260.65 & $-$5.087 & \\
OH     & 16352.22 & 16351.60-16352.80 & $-$4.835 & \\
OH     & 16534.58 & 16534.00-16535.15 & $-$4.746 & \\
OH     & 16538.59 & 16538.00-16539.20 & $-$4.746 & \\
OH     & 16704.36 & 16703.80-16705.00 & $-$4.732 & Blended with two OH lines: \\
       &          &                   &          & 16703.88, log($gf$)=$-$5.383 and 16704.70, log($gf$)=$-$5.383 \\
OH     & 16872.28 & 16871.35-16872.80 & $-$4.975 & Blended with one OH line: 16871.90, log($gf$)=$-$4.999\\
OH     & 16895.18 & 16894.60-16895.80 & $-$4.685 & \\
OH     & 16909.29 & 16908.70-16909.95 & $-$4.654 & \\
\enddata
\end{deluxetable*}

\begin{deluxetable*}{l c|c|ccc|ccc|ccc}[h!]
 \label{tab:results1} 
\tablecaption{The chemical abundances of carbon (using CO lines) and oxygen (using OH lines) and their sensitivity to the variation of  physical parameters  \Teff,  [M/H], and  log($g$) for the host star DH Tau A}  
\tablewidth{0pt}
\tabletypesize{\scriptsize}
\tablehead{
\colhead{Species} & \multicolumn{1}{|c|}{\textit{N}} &  {[X/H]} &
\multicolumn{3}{c|}{${\Delta}\Teff$} & \multicolumn{3}{c|}{$\Delta$[M/H]} & \multicolumn{3}{c}{$\Delta$log($g$)}\\
\colhead{} & \multicolumn{1}{|c|}{}
{} & 
{} & 
 {$-$30} & {+30} & $\overline{\rm{({\Delta}abund)_{T}}}$ &  {$-$0.10} & {+0.10} & $\overline{\rm{(({\Delta}abund)_{M}}}$ &  {$-$0.05} & {+0.05} & $\overline{\rm{({\Delta}abund)_{G}}}$\\
\colhead{} & \multicolumn{1}{|c|}{} &  {} & {(K)} & {(K)}  & {(K)} & {(dex)} & {(dex)} & {(dex)} &  {(dex)} & {(dex)} & {(dex)}
}
\startdata
C (CO) & \multicolumn{1}{|c|}{8} & +0.064  & +0.011 &  $-$0.011 & 0.011 & $-$0.003 & +0.005 & 0.004 & $-$0.040 & +0.047 & 0.044  \\
O (OH) & \multicolumn{1}{|c|}{24} & +0.050 & +0.007 &  $-$0.001 & 0.004 & $-$0.015 & +0.033 & 0.024 & $-$0.028 & +0.041 & 0.035 \\
\enddata
\tablecomments{\textbf{The units are associated with the respective deviated parameters shown in the first row.}}
\end{deluxetable*}

\begin{deluxetable*}{l|ccc|ccc|c|c|c}[h!]
\label{tab:results2} 
\tablecaption{In continuation of Table 3,  the sensitivity of carbon and oxygen abundances to the variation of physical parameters  $\rm{{V_{rot}}}\sin \textit{i}$ and R  as well as the  systematic $\sigma_{\rm{sys}}$, random $\sigma_{\rm{ran}}$, and total uncertainties $\sigma_{\rm{tot}}$ for the host star DH Tau A}  
\tablewidth{0pt}
\tabletypesize{\scriptsize}
\tablehead{
{Species} &  \multicolumn{3}{c|}{$\rm{\Delta{V_{rot}}}\sin \textit{i}$} &  \multicolumn{3}{c|}{$\Delta$R}  & $\sigma_{\rm{sys}}$ & $\sigma_{\rm{ran}}=\rm{std}/\sqrt{N}$ & $\sigma_{\rm{tot}}$    \\
{} & {$-$0.1} & {+0.1} & $\overline{\rm{({\Delta}abund)_{V}}}$ &  {$-$0.03} &  {+0.03} & $\overline{\rm{({\Delta}abund)_{R}}}$ & { } & { } & { } \\
{} &  {km s$^{-1}$} & { km s$^{-1}$} & { km s$^{-1}$} & { } & { } & { } & {} & {} & {} 
}
\startdata
C (CO) & {$-$0.056} & +0.008 & 0.032 & $-$0.037 & +0.036 & 0.037 & 0.067 & 0.041 & 0.079  \\
O (OH) & {$-$0.040} & +0.010 & 0.025 & $-$0.030 & +0.040 & 0.035 & 0.061 & 0.012 & 0.062  \\
\enddata
\tablecomments{\textbf{The units are associated with the respective deviated parameters shown in the first row.}}
\end{deluxetable*}

\section{Discussion: Host-Companion Connection}\label{sec:discussion}
Our measurement of $\rm C/O=0.555\pm0.063$ for DH Tau A is fully consistent with the value for the companion DH TAu b, $\rm C/O=0.54^{+0.06}_{-0.05}$ \citep{Xuan2024b}. This directly validates the chemical homogeneity in this system, which was already suggested in \citet{Xuan2024b} based on previous measurements of Fe abundances for other stars in the Taurus star-forming region \citep{Santos2008, DOrazi2011}. Our results indicate that the $12\pm4~\Mj$ companion is compatible with forming via direct gravitational collapse, either from a disk or a molecular cloud. While formation via core accretion beyond the CO ice line also results in stellar C/O and metallicity as long as the companion accreted more solids than gas \citep{Chachan2023, Xuan2024b}, this scenario is unlikely for DH Tau b due to various reasons. First, core accretion is expected to take several Myr to complete, but the DH Tau system is extremely young with an estimated age of $0.7^{+0.3}_{-0.1}$ Myr from isochrone fitting \citep{Xuan2024b}. Second, the mass ratio between DH Tau b and DH Tau A is $\approx0.026$, which implies that a massive disk is required to form the companion. Massive disks with large host-to-companion mass ratios are prone to gravitational instability, which again makes the core accretion formation scenario unlikely. 

We note that \citet{Xuan2024b} also detected $^{13}$CO in DH Tau b, and found $\co=53^{+50}_{-24}$. In addition to the abundance ratio C/O, future studies should attempt to measure $\co$ for the host stars of giant exoplanets \citep[e.g.][]{Coria_2024, Xuan2024b}, which will allow a direct comparison with the abundance profiles of their companions. However, the fast rotation and veiling of these young host stars may obscure minor isotopologue signatures from carbon- and oxygen-bearing molecules (e.g. H$_2^{18}$O, $^{18}$OH, $^{13}$CO, C$^{18}$O). 

Nevertheless, there are other elemental abundance ratios in addition to C/O that may act as indicators of core accretion formation in sub-stellar companions. Certain volatile abundance ratios in the atmospheres of these sub-stellar objects, when comparable to their host star values, would be indicative of their formation via gravitational instability. When C/N, N/O, S/N, C/S, or O/S ratios deviate significantly from that of their host star, this is a counter-indication of formation via core accretion and subsequent gas/pebble accretion as the planet migrates throughout the disk \citep{Turrini_2021, Pacetti_2022, Crossfield_2023}. With recent detections of H$_2$O, CO, CO$_2$, SO$_2$, H$_2$S, CH$_4$, and NH$_3$ in brown dwarfs and gas giant exoplanets \citep{Schlawin_2024, Hsu2024, Fu_2024, Rafi_2024, Yang_2024, Powell_2024, Nasedkin2024, Barrado_2023}, these volatile abundance ratios may be used for direct comparisons when measured in both the host star and in the sub-stellar companion and provide a better insight into their formation scenarios.

\section{Summary}\label{sec:summary}
We measure the O and C abundances as well as the C/O abundance ratio of the young M-type star DH Tau A using its high-resolution IGRINS spectra ($\sim$ 45000) in both the $H$ and $K$ band. We determine the star's physical parameters (Table \ref{tab:parameters}), which are then used as input for our abundance measurements. By a careful visual inspection,  we select 24 OH (in the $H$ band) and 8 CO (in the $K$ band) lines (Table \ref{tab:line_data}) to infer the elemental abundances of O and H, respectively.  We briefly describe our approach to normalize the observed spectrum by illustrating two examples (Figures \ref{fig:OH_normalization} and \ref{fig:CO_normalization}). We finally apply our automatic code, AutoSpecFit, which performs an iterative, line-by-line ${\chi}^{2}$ minimization over all the selected lines until the final abundances of O and C are reached simultaneously (Table \ref{tab:results1}). 

We find a near-solar $\rm C/O=0.555\pm0.063$ for the host star, which is completely consistent with that of the companion DH Tau b, $\rm C/O=0.54^{+0.06}_{-0.05}$. This confirms that  the two components are chemically homogeneous, which suggests direct gravitational collapse (with a timescale of $<$1 Myr) as a more probable mechanism for the formation of the companion. Given the very young age of the system ($0.7^{+0.3}_{-0.1}$ Myr), it is unlikely that the companion was formed through a core accretion process (with a typical time scale of several Myr). 

The comparison between the chemical abundances of planets and substellar objects with those of their host stars can provide essential clues on the formation pathways of  host-companion systems. With the advent of space-based telescopes such as JWST, chemical composition measurements for substellar companions have become possible. While the chemical abundances of many hotter JWST FGK-dwarf hosts  have already been measured  \citep[e.g.][]{Kolecki_Wang2022, Polanski2022}, due to the complexity of M-dwarf spectra, only a small number of cooler JWST M-dwarf hosts have reported abundances \citep[e.g.][]{Hejazi2024, Melo2024}. Spectroscopic analyses of M-type host stars, such as the one presented in this paper, are paramount to understanding the condition of protoplanetary disks and the subsequent formation of planets or low-mass substellar companions. This work has demonstrated successful near-infrared spectral fitting of an actively accreting M-dwarf star and will allow for the derivation of high-precision, multi-element abundances of several high-priority JWST M-dwarf exoplanet hosts and hosts to young, directly imaged exoplanets and brown dwarfs including the targets in \cite{Xuan2024a} and many others.

\

\noindent
{Acknowledgements:} We wish to thank the anonymous referee for their insightful comments and suggestions, which improved our manuscript. We express our appreciation to Justin Cantrell for his technical support with the high-performance computing system of the physics and astronomy department, Georgia State University, which was used for this study. N.H. and I.J.M.C. acknowledge support from NSF AAG grant No. 2108686 and from NASA ICAR grant No. NNH19ZDA001N. J.X. is supported by the NASA Future Investigators in NASA Earth and Space Science and Technology (FINESST) award \#80NSSC23K1434. 

This work used The Immersion Grating Infrared Spectrometer (IGRINS) was developed under a collaboration between the University of Texas at Austin and the Korea Astronomy and Space Science Institute (KASI) with the financial support of the US National Science Foundation under grants AST-1229522, AST-1702267 and AST-1908892, McDonald Observatory of the University of Texas at Austin, the Korean GMT Project of KASI, the Mt. Cuba Astronomical Foundation and Gemini Observatory. This paper includes data taken at The McDonald Observatory of The University of Texas at Austin.

\pagebreak


\pagebreak


\end{document}